\documentclass[12pt]{article}
\def\be{\begin{equation}}
\def\ee{\end{equation}}
\def\bea{\begin{eqnarray}}
\def\eea{\end{eqnarray}}
\usepackage{graphicx}% Include figure files

\catcode`\@=11
\def\lsim{\mathrel{\mathpalette\@versim<}}
\def\gsim{\mathrel{\mathpalette\@versim>}}
\def\@versim#1#2{\vcenter{\offinterlineskip
\ialign{$\m@th#1\hfil##\hfil$\crcr#2\crcr\sim\crcr } }}
\catcode`\@=12

\catcode`@=12
\begin{document}

\thispagestyle{empty}
\begin{flushright}
UCRHEP-T509\\
August 2011\
\end{flushright}
\vspace{0.3in}
\begin{center}
{\LARGE \bf Exceeding the MSSM Higgs Mass Bound\\ 
in a Special Class of U(1) Gauge Models\\}
\vspace{1.5in}
{\bf Ernest Ma\\}
\vspace{0.2in}
{\sl Department of Physics and Astronomy, University of California,\\ 
Riverside, California 92521, USA\\}
\end{center}
\vspace{1.5in}
\begin{abstract}\
A special class of supersymmetric U(1) gauge extensions of the standard model 
was proposed in 2002.  It is anomaly-free, has no $\mu$ term, and conserves 
baryon and lepton numbers automatically.  It also allows the lightest Higgs 
boson to have a mass exceeding the MSSM (Minimal Supersymmetric Standard Model) 
bound, i.e. about 130 GeV, which is of current topical interest from LHC 
(Large Hadron Collider) data.  This and other new aspects of this 2002 
proposal are discussed.
\end{abstract}

\newpage
\baselineskip 24pt

Supersymmetry is a very attractive theoretical idea, but as the standard model 
of particle interactions is extended to include supersymmetry, several new 
problems arise.

\begin{itemize}

\item{
(1) Whereas the particle content of the $SU(3)_C \times 
SU(2)_L \times U(1)_Y$ gauge group of the standard model (SM) guarantees 
automatically the conservation of both baryon number $B$ and lepton number 
$L$, as far as renormalizable interactions are concerned, the superfield 
content of its supersymmetric extension allows trilinear 
terms in its superpotential which violate both $B$ and $L$.  As a result, 
the proton would decay immediately, rendering impossible our existence. 
Even with the conventional imposition of $R$ parity, 
i.e.~$R \equiv (-1)^{2j+3B+L}$, which forbids these terms, effective 
dimension-five operators still exist in supersymmetry which mediate 
proton decay, whereas in the standard model, such operators are dimension-six.  
}

\item{
(2) The $\mu \hat{\phi}_1 \hat{\phi}_2$ term in the superpotential of the two 
Higgs superfields $\hat{\phi}_{1,2}$ is invariant under supersymmetry, hence 
there is no understanding as to why the value of $\mu$ is also close to 
(and not much higher than) the presumed scale of supersymmetry breaking, 
i.e.~1 TeV or so, which is required for supersymmetry to be relevant in  
solving the hierarchy problem of the standard model and the rationale for 
adopting it in the first place.
}

\item{
(3) Supersymmetry imposes an upper bound on the mass of the lightest physical 
Higgs scalar boson $H_1^0$ of about 130 GeV~\cite{d08}, which is reached in 
the limit of large $\tan \beta = v_2/v_1 = \langle \phi_2^0 \rangle/
\langle \phi_1^0 \rangle$.  In the decoupling limit, $H_1^0 = \phi_1^0 
\cos \beta + \phi_2^0 \sin \beta$ is essentially identical to the one 
physical Higgs boson $H^0$ of the standard model.  The current experimental 
lower bound~\cite{pdg10} on the mass of $H^0$ is 114.4 GeV from LEP2, with 
the exclusion of 158 to 175 GeV from the Tevatron.  However, recent LHC 
data~\cite{lhc11} have excluded most of the mass region above 150 GeV to 450 
GeV.   If there is no Higgs boson below 130 GeV, the MSSM 
(Minimal Supersymmetric Standard Model) would be excluded, unless the 
supersymmetry breaking scale is much greater than 1 TeV.
}

\end{itemize}

In 2002, a remarkable class of supersymmetric $U(1)_X$ anomaly-free 
gauge extensions of the standard model was discovered~\cite{m02}, with 
particle content given in Table 1. It addresses all three of the above issues.
\begin{table}[htb]
\begin{center}
\begin{tabular}{|c|c|c|c|c|}
\hline
superfield & copies & $SU(3)_C \times SU(2)_L \times U(1)_Y$ & $U(1)_X - (A)$ & 
$U(1)_X - (B)$ \\ 
\hline
$\hat{Q}=(\hat{u}, \hat{d})$ & 3 & $(3,2,1/6)$ & $n_1$ & $n_1$ \\ 
$\hat{u}^c$ & 3 & $(3^*,1,-2/3)$ & $(7n_1 + 3n_4)/2$ & $5n_1$ \\ 
$\hat{d}^c$ & 3 & $(3^*,1,1/3)$ & $(7n_1 + 3n_4)/2$ & $2n_1 + 3n_4$ \\ 
\hline
$\hat{L}=(\hat{\nu}, \hat{e})$ & 3 & $(1,2,-1/2)$ & $n_4$ & $n_4$ \\ 
$\hat{e}^c$ & 3 & $(1,1,1)$ & $(9n_1 + n_4)/2$ & $3n_1 + 2n_4$ \\ 
$\hat{N}^c$ & 3 & $(1,1,0)$ & $(9n_1 + n_4)/2$ & $6n_1 - n_4$ \\ 
\hline
$\hat{\phi}_1=(\hat{\phi}_1^0, \hat{\phi}_1^-)$ & 1 & $(1,2,-1/2)$ & 
$-(9n_1 + 3n_4)/2$ & $-3n_1 - 3n_4$ \\ 
$\hat{\phi}_2=(\hat{\phi}_2^+, \hat{\phi}_2^0)$ & 1 & $(1,2,1/2)$ & 
$-(9n_1 + 3n_4)/2$ & $-6n_1$ \\ 
$\hat{S}_3$ & 2 & $(1,1,0)$ & $9n_1 + 3n_4$ & $9n_1 + 3n_4$ \\ 
\hline
$\hat{U}$ & 2 & $(3,1,2/3)$ & $-4n_1 - 2n_4$ & $-6n_1$ \\ 
$\hat{D}$ & 1 & $(3,1,-1/3)$ & $-4n_1 - 2n_4$ & $-6n_1$ \\ 
$\hat{U}^c$ & 2 & $(3^*,1,-2/3)$ & $-5n_1 - n_4$ & $-3n_1 - 3n_4$ \\ 
$\hat{D}^c$ & 1 & $(3^*,1,1/3)$ & $-5n_1 - n_4$ & $-3n_1 - 3n_4$ \\ 
\hline
$\hat{S}_2$ & 3 & $(1,1,0)$ & $-6n_1 - 2n_4$ & $-6n_1 - 2n_4$ \\ 
$\hat{S}_1$ & 3 & $(1,1,0)$ & $-3n_1 - n_4$ & $-3n_1 - n_4$ \\ 
\hline
\end{tabular}
\caption{$U(1)_X$ charges for all the superfields of this anomaly-free 
supersymmetric model.  There are two possible solutions: (A) and (B).}
\end{center}
\end{table}

\begin{itemize}

\item{
(1) As in the standard model, both $B$ and $L$ are separately conserved 
automatically, as far as renormalizable interactions are concerned, 
and there are no dimension-five operators which mediate proton 
decay~\cite{dlt06}.  Dimension-six operators exist if $S_1$ has a 
vacuum expectation value.
}

\item{
(2) The $\mu \hat{\phi}_1 \hat{\phi}_2$ term is replaced by 
$f \hat{S}_3 \hat{\phi}_1 \hat{\phi}_2$, where the singlet superfield 
$\hat{S}_3$ transforms nontrivially 
under $U(1)_X$.  The soft breaking of supersymmetry also breaks $U(1)_X$, 
so that the two scales are naturally the same.  Note that in many 
supersymmetric gauge $U(1)$ extensions of the standard model, such as 
those from $E_6$, which also eliminate the $\mu$ term, $B$ and $L$ are 
in general not conserved, without the imposition of extra symmetries 
as in the MSSM.  Here $U(1)_X$ by itself is sufficient for this purpose. 
Note also that $U(1)_{B-L}$ cannot eliminate the $\mu$ term.
}

\item{
(3) The MSSM bound of 130 GeV on the mass of $H^0_1$ is allowed to be 
exceeded, because the Higgs sector now includes a singlet~\cite{kkw93}.  
For a specific $U(1)$ gauge factor, there is still a bound,  
which has been discussed over the 
years~\cite{dm93,mn94,dm95,km96,cdeel97,blls06}.  However, for most known 
supersymmetric $U(1)$ models~\cite{km97,mm11}, 144 GeV is still difficult 
to reach.  (This value of 144 GeV is chosen for illustration.  It can be 
replaced by any value above 135 GeV or so.)  In this special class of 
$U(1)_X$ models, it will be shown that the $H_1^0$ mass is allowed to be 
144 GeV, while keeping the supersymmetry breaking scale at 1 TeV.  It also 
has verifiable predictions at the LHC in terms of exotic new quarks.
}

\end{itemize}

In the above, all superfields are left-handed, with the usual right-handed 
fields represented by their charge conjugates. There are three copies each 
of $\hat{Q}, \hat{u}^c, \hat{d}^c, \hat{L}, \hat{e}^c, \hat{N}^c, \hat{S}_2, 
\hat{S}_1$, two copies each of $\hat{U}, \hat{U}^c, \hat{S}_3$, and one copy 
each of $\hat{\phi}_1, \hat{\phi}_2, \hat{D}, \hat{D}^c$.  The resulting field 
theory is free of all three possible quantum anomalies, i.e. axial-vector, 
global SU(2), and mixed gravitational-gauge, for all values of $n_1$ and $n_4$ 
in both solutions (A) and (B)~\cite{m02}.  The allowed interactions in all 
cases are
\begin{eqnarray}
\hat{Q} \hat{u}^c \hat{\phi}_2, ~~ \hat{Q} \hat{d}^c \hat{\phi}_1, 
~~ \hat{L} \hat{e}^c \hat{\phi}_1, ~~ \hat{L} \hat{N}^c \hat{\phi}_2,  
~~ \hat{S}_3 \hat{\phi}_1 \hat{\phi}_2, ~~  \hat{S}_3 \hat{U} \hat{U}^c, ~~  
\hat{S}_3 \hat{D} \hat{D}^c, ~~  \hat{S}_3 \hat{S}_1 \hat{S}_2.   
\end{eqnarray}
The $U(1)_X$ gauge symmetry is broken by $\langle S_3 \rangle$ for 
$3n_1+n_4 \neq 0$, through which the particles contained in $\hat{U},
\hat{U}^c,\hat{D},\hat{D}^c,\hat{S}_{1,2,3}$ obtain their large masses. 
At the same time, an effective $\mu \hat{\phi}_1 \hat{\phi}_2$ term is 
generated.

Phenomenologically, the exotic new quarks must decay.  Assuming that $u^c$ 
is distinguished from $U^c$ and $d^c$ from $D^c$ by $U(1)_X$ (otherwise the 
$3 \times 3$ quark mixing matrix would not be unitary), this requirement 
leads to two possible models if only renormalizable interactions are 
considered.  A third possible model is shown as an example if 
higher-dimensional operators are included.  There may be other viable 
examples of this kind.

\begin{itemize}

\item{
(1) As already discussed in Ref.~\cite{m10}, $n_1=0$ in Solution (A) implies 
the following additional interactions:
\begin{equation}
\hat{u}^c \hat{e}^c \hat{D}, ~~ \hat{u}^c \hat{N}^c \hat{U}, ~~ 
\hat{d}^c \hat{N}^c \hat{D}, ~~ \hat{Q} \hat{L} \hat{D}^c, ~~ 
\hat{N}^c \hat{N}^c \hat{S}_1. 
\end{equation}
This means that scalar $U,D$ are leptoquarks and their observation at the LHC 
is assured if kinematically allowed.  The $\hat{N}^c \hat{N}^c \hat{S}_1$ term 
also means that lepton number $L$ becomes multiplicative, i.e. 
$(-1)^L$, and the neutrinos obtain small Majorana masses from the seesaw 
mechanism as $\langle S_{1,2} \rangle$ also acquire vacuum expectation values.
}

\item{
(2) $n_4=-n_1$ in Solution (B) implies the following additional interactions:
\begin{equation}
\hat{Q} \hat{L} \hat{D}^c, ~~ \hat{u}^c \hat{e}^c \hat{D}, ~~ 
\hat{d}^c \hat{N}^c \hat{D}, ~~ \hat{U}^c \hat{D}^c \hat{D}^c. 
\end{equation}
This means that scalar $D$ is a leptoquark and scalar $U^c$ is a dileptoquark. 
However, $\hat{D}^c \hat{D}^c$ in the term $\hat{U}^c \hat{D}^c \hat{D}^c$ 
must transform as $\underline{3}$ under $SU(3)_C$, i.e. antisymmetric, which 
is impossible for one copy of $\hat{D}^c$.  To allow this term, a second pair 
of $\hat{D}$ and $\hat{D}^c$ should be added, with the new $\hat{D}^c$ 
transforming as the existing $\hat{D}^c$, i.e. trivially under $U(1)_X$, 
and the new $\hat{D}$ also trivially, i.e. differently from the existing 
$\hat{D}$, so that no new anomaly is generated.  Neutrino masses come 
only from $\langle \phi_2^0 \rangle$, hence they are Dirac.
}

\item{
(3) $n_4=n_1$ leads to the same model in both Solutions (A) and (B). 
However, no additional renormalizable interaction is implied.  As such, 
the exotic quarks appear to be stable.  Nevertheless, the higher-dimensional 
operators
\begin{equation}
\hat{d}^c \hat{d}^c \hat{U}^c \hat{S}_1, ~~ \hat{d}^c \hat{u}^c \hat{D}^c 
\hat{S}_1, ~~  \hat{u}^c \hat{N}^c \hat{U} \hat{S}_1, ~~\hat{d}^c \hat{N}^c 
\hat{D} \hat{S}_1, 
\end{equation}
are allowed, but suppressed by a large mass scale, such as the 
grand-unification scale of perhaps $10^{16}$ GeV or the Planck scale of 
$10^{19}$ GeV.  The exotic quarks $U$ and $D$ now have $B-L=-2/3$ and 
would decay into quarks and leptons through a dimension-five operator 
if $S_1$ has a vacuum expectation value as in (1). Here neutrino masses 
come only from $\langle \phi_2^0 \rangle$, i.e. Dirac as in (2).

As $S_1$ picks up a nonzero vacuum expectation value at the TeV scale, 
$B-L$ is still conserved, but not $B$ and $L$ separately.  this means that 
proton decay via $p \to \pi^+ \bar{\nu}$ is possible, but it comes from a 
dimension-six operator as in the SM, suppressed by two powers of the 
large mass scale implied by Eq.(4). 
}

\end{itemize}

All of the above cases forbid the terms
\begin{equation}
\hat{L} \hat{e}^c \hat{L}, ~~ \hat{Q} \hat{d}^c \hat{L}, ~~ 
\hat{u}^c \hat{d}^c \hat{d}^c, ~~ \hat{L} \hat{\phi}_2, 
\end{equation}
which are otherwise allowed in the MSSM without the imposition of $R$ parity. 
Thus $B$ and $L$ are conserved by the renormalizable interactions of this 
theory.  The lowest-order higher-dimensional operator for proton decay is 
$\hat{Q} \hat{Q} \hat{Q} \hat{L} \hat{S}_1$, which is dimension-six if 
$S_1$ has a vacuum expectation value as in (1) and (3). 

Let the $U(1)_X$ gauge coupling $g_X$ be normalized by defining the charge 
of $S_3$, i.e. $9n_1 + 3n_4$, to be unity.  Consider then the production of 
$X$ at the LHC and its decay branching fraction into $\mu^- \mu^+$. 
Compare these to the case of a hypothetical $Z'$ of the same mass, but
with couplings to quarks and leptons as in the standard model.  Their 
relative factors are given in Table 2.
\begin{table}[htb]
\begin{center}
\begin{tabular}{|c|c|c|c|c|}
\hline
Model & Production & $B(\mu^- \mu^+)$ & Event Ratio & Max($g_X^2$) \\ 
\hline
SM & 0.47 & 0.03 & 1 & --- \\
\hline
(1) & 3/4 & 5/189 & 1.41 $(g_X^2/g_Z^2)$ & 0.394 \\
(2) & 3/2 & 2/261 & 0.82 $(g_X^2/g_Z^2)$ & 0.680 \\ 
(3) & 13/24 & 26/549 & 1.82 $(g_X^2/g_Z^2)$ & 0.305 \\ 
\hline 
\end{tabular}
\caption{Relative factors of production and decay at the LHC for a 
standard-model $Z'$ and the $X$ boson in models (1),(2),(3). The maximum 
allowed value of $g_X^2$ is obtained if $m_X = 2$ TeV is assumed.}
\end{center}
\end{table}

The production is assumed to be proportional to $2[(g_L^u)^2 + (g_R^u)^2] + 
(g_L^d)^2 + (g_R^d)^2$, and the decay branching fractions of $Z'$ and $X$ to 
$\mu^- \mu^+$ assumes that their only decay products are the usual 
leptons and quarks, including $t \bar{t}$ which is of course not possible 
for the actual $Z$ boson at 91.2 GeV.   The event ratio for observing 
$\mu^- \mu^+$ relative to the standard model is in units of $g_X^2/g_Z^2$.  
From the nonobservation of a standard-model $Z'$ below about 2 TeV at the 
LHC~\cite{c11,t11}, an upper bound on $g_X^2$ is obtained if $m_X = 2$ TeV 
is assumed.  These bounds are relaxed if $X$ decays into particles other 
than the usual leptons and quarks, or if $m_X > 2$ TeV.

The effective two-Higgs-doublet potential of any supersymmetric $U(1)_X$ 
gauge extension of the standard model is given by
\begin{eqnarray}
V &=& m_1^2 \phi_1^\dagger \phi_1 + m_2^2 \phi_2^\dagger \phi_2 + m_{12}^2 
(\phi_1^\dagger \phi_2 + \phi_2^\dagger \phi_1) \nonumber \\ 
&+& {1 \over 2} \lambda_1  (\phi_1^\dagger \phi_1)^2 + 
{1 \over 2} \lambda_2  (\phi_2^\dagger \phi_2)^2 + 
\lambda_3  (\phi_1^\dagger \phi_1)(\phi_2^\dagger \phi_2) + 
\lambda_4  (\phi_1^\dagger \phi_2)(\phi_2^\dagger \phi_1),
\end{eqnarray}
where~\cite{km97} 
\begin{eqnarray}
\lambda_1 &=& {1 \over 4} (g_1^2 + g_2^2) + 2af^2 - {f^4 \over g_X^2}, \\ 
\lambda_2 &=& {1 \over 4} (g_1^2 + g_2^2) + 2(1-a)f^2 - {f^4 \over g_X^2}, \\ 
\lambda_3 &=& -{1 \over 4} g_1^2 + {1 \over 4} g_2^2 + f^2 - {f^4 \over g_X^2}, 
\\ 
\lambda_4 &=& -{1 \over 2} g_2^2 + f^2.
\end{eqnarray}
In the above, the contributions of the soft supersymmetry breaking scalar 
trilinear term $f A_f S_3 \phi_1 \phi_2$ are assumed to be small, i.e. 
$f A_f / g_X^2 \langle S_3 \rangle << 1$.  Also $\phi_1$ in $V$ has been 
redefined from $(\phi_1^0,\phi_1^-)$ to $(\phi_1^+,\phi_1^0)$ to agree with 
the usual convention for analyzing two Higgs doublets~\cite{dm78}. The 
$U(1)_X$ charges of $\phi_{1,2}$ in $V$ are $a$ and $a-1$.  The upper bound 
on the $H_1^0$ mass is then given by~\cite{km97}
\begin{equation}
m^2(H_1^0) < M_Z^2 \cos^2 2 \beta + \epsilon + f^2v^2 [3 + 2(2a-1) \cos 2 \beta 
- \cos^2 2 \beta - 2 f^2/g_X^2],
\end{equation}
where $v = 174$ GeV and~\cite{d08}
\begin{equation}
\epsilon = {3 g_2^2 m_t^4 \over 8 \pi^2 M_W^2} \ln \left( 1 + {\tilde{m}_{eff}^2 
\over m_t^2} \right)
\end{equation}
is the well-known radiative correction due to the $t$ quark and its scalar 
counterparts which are represented by $\tilde{m}_{eff}$, usually set equal to 
1 TeV.  For a given value of $g_X^2$, this bound is maximized by 
\begin{equation}
{f^2 \over g_X^2} = {3 \over 4} + \left( a-{1 \over 2} \right) \cos 2 \beta 
- {1 \over 4} \cos^2 2 \beta,
\end{equation}
resulting in
\begin{equation}
m^2(H_1^0) < M_Z^2 \cos^2 2 \beta + \epsilon + {1 \over 8} g_X^2 v^2 
[3 + 2(2a-1) \cos 2 \beta - \cos^2 2 \beta]^2.
\end{equation}

Now $a=1/2$ in both models (1) and (3). The bound 
on $m^2(H_1^0)$ is maximized for $\cos^2 2 \beta = 1$, i.e.
\begin{equation}
m^2(H_1^0) < M_Z^2 + \epsilon + {1 \over 2} g_X^2 v^2,
\end{equation}
which is consistent with $m(H_1^0) = 144$ GeV if $g_X^2 > 0.292$. 
According to Table 2, this is allowed by both models.  On 
the other hand, the $Z-X$ mixing angle is given by~\cite{km97}
\begin{equation}
\theta_{ZX} \simeq - {g_Z g_X (\sin \beta - a) v^2 \over m_X^2},
\end{equation}
which must be very small, say $5 \times 10^{-4}$.  For $\sin^2 \beta = 1$ 
and $a = 1/2$, $g_X^2 = 0.292$, $m_X > 3.5$ TeV is required.  This means 
that the bounds in Table 2 are relaxed for both models.

If $\sin^2 \beta = 1/2$, there is no $Z-X$ mixing and $m_X$ is not 
constrained.  Now
\begin{equation}
m^2(H_1^0) < \epsilon + {9 \over 8} g_X^2 v^2,
\end{equation}
and $m(H_1^0) = 144 $ GeV requires $g_X^2 > 0.374$ which is still possible in 
model (1).

For model (2), $a=0$. Now
\begin{equation}
m^2(H_1^0) < M_Z^2 \cos^2 2 \beta + \epsilon + {1 \over 8} g_X^2 v^2 
(1 - \cos 2 \beta)^2(3 + \cos 2 \beta)^2.
\end{equation}
If $\tan \beta$ is large, 
\begin{equation}
m^2(H_1^0) < \epsilon + 2 g_X^2 v^2.
\end{equation}
This is consistent with $m(H_1^0) = 144$ GeV if $g_X^2 > 0.073$, which is 
allowed by Table 2.  However, $Z-X$ mixing requires $m_X > 3.9$ TeV.

In summary, it has been shown that $m(H_1^0) = 144$ GeV is possible in all 
three $U(1)_X$ models (the first of which was considered earlier~\cite{m10}), 
with different assumptions on the values of $\cos 2 \beta$ 
and $m_X$ in each case.  This would be especially important if experimental 
data rule out a Higgs boson below the MSSM bound of about 130 GeV.  Each 
model is also associated with different predictions of how the exotic $U$ 
and $D$ quarks and scalar quarks decay. As more data accumulate at the LHC, 
these ideas will be tested, and be rejected or reinforced.

This work is supported in part by the U.~S.~Department of Energy under Grant 
No.~DE-AC02-06CH11357.

%\newpage
\baselineskip 16pt
\bibliographystyle{unsrt}

\end{document}